# Large electric-field induced strain in $BiFeO_3$ ceramics

Tadej Rojac,[1,a)] Marija Kosec,[1] Dragan Damjanovic[2]

[1] Jožef Stefan Institute, Jamova cesta 39, 1000 Ljubljana, Slovenia

[2] Swiss Federal Institute of Technology in Lausanne - EPFL, Ceramics Laboratory, Lausanne 1015, Switzerland

Large bipolar strain of up to 0.36% (peak-to-peak value) was measured in $BiFeO_3$ ceramics at low frequency (0.1 Hz) and large amplitude (140 kV/cm) of the driving field. This strain is comparable to that achievable in highly efficient Pb-based perovskite ceramics, such as $Pb(Zr,Ti)O_3$ and $Pb(Mg,Nb)O_3$–$PbTiO_3$. The strain showed a strong dependence on the field frequency and is likely largely associated with domain switching involving predominantly non-180° domain walls. In addition, rearrangement of charged defects by applying electric field of low frequency depins these domain walls, resulting in a more efficient switching and, consequently, an increased response.

[a)] Electronic mail: tadej,rojac@ijs.si

Bismuth ferrite ($BiFeO_3$) has recently been subject of intensive research, driven primarily by its ability to exhibit both ferroelectric and antiferromagnetic ordering.[1,2] Significant efforts have been put particularly in understanding of how to manipulate magnetic ordering by applying electric field (magnetoelectric coupling). Since this coupling requires ferroelastic rather than ferroelectric domain reversal,[3] the switching of ferroelectric-ferroelastic 71° and 109° domains in this rhombohedral structure, which is accompanied by strain in the material, has been of particular interest.[4–6] Strain–electric-field relationship in epitaxial $BiFeO_3$ thin films has been recently reported by Zeches et al.[7] and Zhang et al.[8] Bipolar strain as large as 5% was measured at fields of >1000 kV/cm and was associated with phase transformation (motion of the interphase boundary). On the other hand, maximum strain of 0.2% was reported in a $BiFeO_3$ film prepared by a chemical method.[9] Even smaller strain was found in $BiFeO_3$ ceramics (0.07%), which was, however, measured at 60 kV/cm, i.e., well below the coercive field of 100 kV/cm.[10] Application of large fields in bulk ceramics of $BiFeO_3$ is challenging, especially at low frequencies, due to the high conductivity. Since bipolar strain larger than presently reported can be expected in $BiFeO_3$ ceramics, further studies on the strain versus electric-field relationship are needed.

Exploiting the possibility of applying high electric fields of low frequency, we report here a large electric-field induced strain in $BiFeO_3$ ceramics. A tentative explanation is proposed for its origin.

$BiFeO_3$ ceramics were prepared by sintering a mechanochemically-activated $Bi_2O_3$–$Fe_2O_3$ powder mixture at 760°C for 6 hours. The synthesis procedure is described elsewhere.[10] X-ray diffraction analysis (PANalytical X'Pert Pro diffractometer) of the ceramics showed phase-pure $BiFeO_3$, however, a small amount of secondary phases, rich in Bi and Fe, was detected using scanning electron microscopy (SEM, JEOL JSM-7600F). The concentration of these phases, as estimated from an SEM image,[11] was around 3% (in area fraction). The relative geometrical density of the pellets was 93% and the grain size, which was determined from an SEM image of a thermally etched sample,[12] was 1.8±0.9 μm. For the electrical measurements the sintered pellets were thinned to approximately 0.2 mm, polished and electrode with Cr/Au by sputtering. Strain and polarization measurements were done simultaneously using an aixACT TF 2000 analyzer. The samples were measured in a "single loop mode", i.e., single sinusoidal electric-field waveforms of defined frequency and amplitude were applied. Due to high voltages, the samples were immersed in a silicone oil. Strain–electric-field curves were plotted by taking the initial strain to be zero.

Strain–electric-field hysteresis loop of BiFeO$_3$ ceramics obtained by applying a single period of the sinusoidal field of amplitude 140 kV/cm at 0.1 Hz is shown in Fig. 1. Distinct features of a typical "butterfly"-shaped hysteresis loop can be observed: after an initial decrease, the strain reaches a minimum at the coercive field of around 90 kV/cm, then makes a steep increase until maximum field and, finally, decreases again to a remanent value as the field is released; same trend is repeated for the field cycle of negative polarity. In analogy with other ferroelectric materials, this behavior is commonly linked to switching and movement of domain walls by electric field, particularly of the non-180° domain walls, which may involve a significant change in the dimensions of the grains in the ceramics.[13–16] The absence of a more linear part in the strain-field relationship at high positive and negative electric fields, such as observed, for example, in Pb(Zr,Ti)O$_3$ (PZT),[14,17] suggests that strain is not saturated and switching is probably not completed even at 140 kV/cm. Experimentally, dielectric breakdown was typically observed above this field.

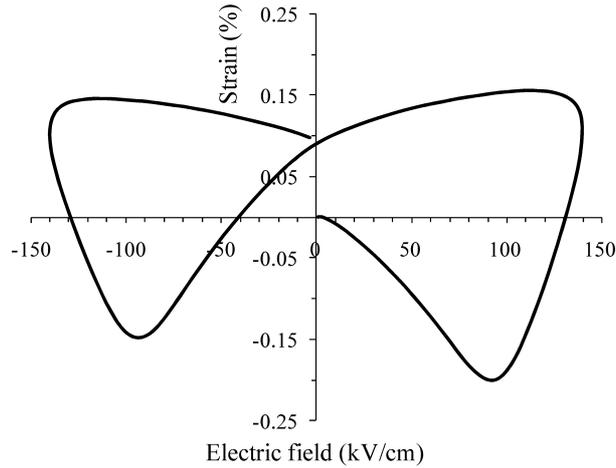

FIG. 1. Strain–electric-field hysteresis loop of BiFeO$_3$ measured at 0.1 Hz and 140 kV/cm of maximum field amplitude.

The most remarkable feature of the hysteresis loop in Fig. 1 is large peak-to-peak strain (difference between maximum and minimum strain), which reaches 0.36%. This value is comparable to the bipolar strain achieved in most known Pb-based perovskite ceramics, such as PZT[14,16–18] and Pb(Mg,Nb)O$_3$–PbTiO$_3$ (PMN-PT),[19,20] as well as in lead-free (Na$_{0.5}$Bi$_{0.5}$)TiO$_3$–BaTiO$_3$– (K,Na)NbO$_3$.[21] In addition, the ratio between peak-to-peak strain and electric field amplitude in our ceramics (0.36% at 140 kV/cm) is comparable to that obtained in epitaxial BiFeO$_3$ thin films (5% at 1500 kV/cm) reported in ref.8. In the following, we limit our discussion to domain reversal as the most probable origin of the measured strain, however, due to large electric field applied, a possibility of electrically driven phase transition, like the one recently demonstrated between tetragonal-like and rhombohedral polymorphs in BiFeO$_3$ epitaxial films,[22] should not be excluded.

It should be stressed that the strain-field curves, similar to the one shown in Fig. 1, were observed on several samples, which were processed in different ways, e.g., directly sintered without calcination or sintered after calcination, and contained different amounts (1–5%) of secondary phases ($Bi_{25}FeO_{39}$ and $Bi_2Fe_4O_9$). Thus, the large strain is little influenced by minor amount of impurity phases and is likely related intrinsically to $BiFeO_3$.

In order to explore in more detail the origin of the large strain, the displacement of the sample was measured by applying electric field of 120 kV/cm of amplitude at different frequencies. The strain-field curves measured at 100, 10, 1 and 0.1 Hz are shown in Fig. 2. Strong frequency dependence is evident, both in the magnitude and qualitative aspect of the response. At the driving field of 100 Hz the sample did not show any measurable displacement up to 100 kV/cm. After an increase of strain to 0.045% above this threshold field, a strong restoring force is observed upon releasing the field, evidenced by zero remanent strain. Note that this behavior is in striking contrast to the response at 0.1 Hz showed in Fig.1.

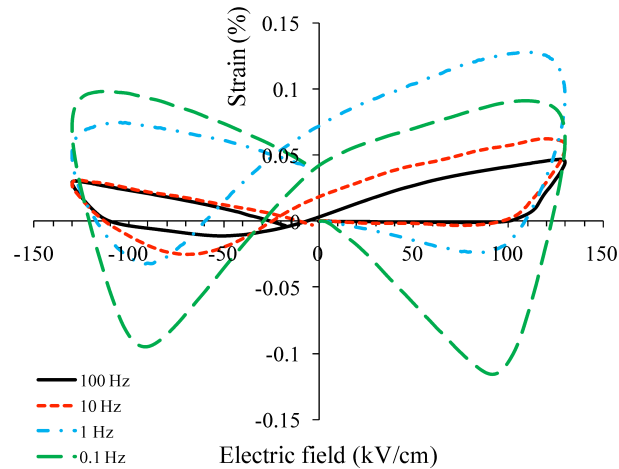

FIG. 2. Strain–electric-field hysteresis loops of $BiFeO_3$ measured at different frequencies and 130 kV/cm of maximum field amplitude.

At 10 Hz similar behavior to the one at 100 Hz is observed (Fig. 2); the main difference is the higher peak-to-peak strain, which is 0.08% at 10 Hz compared to 0.05% at 100 Hz. A more opened, but still asymmetrical, curve is observed at 1 Hz with larger peak-to-peak strain of 0.16%. Finally, an open and a more symmetrical loop is observed at the lowest measuring frequency of 0.1 Hz, which also gave the largest strain (0.21%). We can infer from these results that application of electric field of low frequency enables more efficient switching of domains in BiFeO$_3$ and, consequently, larger strain response. It appears of a crucial importance, therefore, to prepare BiFeO$_3$ ceramics that may sustain electric field of low frequency and large amplitude. In the following, based on our experimental results and observations from the literature, a tentative explanation will be given for the large strain.

Frequency dependence of strain, like the one shown in Fig. 2 for BiFeO$_3$, was observed also in other perovskites, e.g., PZT,[18] PMN-PT[19] and Pb(Zn,Nb)O$_3$–PbTiO$_3$ (PZN-PT).[23] In a single crystal of PZN-PT, such dependence was interpreted as being a consequence of a two-stage non-180° domain switching mechanism, which was confirmed by in-situ neutron diffraction analysis. According to this mechanism, polarization reversal occurs via an intermediate state characterized by large amount of non-180° domains, which then switch for a second time to fully reverse the polarization by 180°. Slow movement of these non-180° domain walls was directly related to the strong frequency dependence of macroscopic strain in PZN-PT, i.e., electric field of low frequency allows kinetically a more efficient switching of these low-mobile domain walls, resulting in larger strain.[23] The same mechanism has also been evidenced in PZT[24] and La-modified PZT ceramics,[25] and was proposed to explain the frequency dependence of the field-induced strain.[18] Slow movement of non-180° domain walls, which is not necessarily related to intermediate switching, has been discussed in refs.[14,26]; it was proposed that the frequency dependence has origin in the strain that the material needs to accommodate during ferroelastic reorientation of domains.

Already in 1990 Kubel and Schmid[27] have predicted that 180° polarization reversal in BiFeO$_3$ is energetically less favorable than reversal by 71° and 109° because the last two reversals require smaller ion displacements. This finding goes in favor of a switching mechanism via

non-180° domains, as discussed above. Such a possibility was also confirmed by Baek et al.[6] using phase-field simulation. Finally, the difficulty of 180° switching was observed experimentally during poling of a BiFeO$_3$ single crystal.[28] Thus, according to observations in the literature, there is a possibility that the large strain observed here in BiFeO$_3$ ceramics at low frequencies (Fig. 1) and the associated characteristic frequency dependence (Fig. 2) are related to switching of predominantly non-180° domains (71° and 109° in rhombohedral system).

Movement and switching of domain walls in "hard" ferroelectric materials, which are strongly restricted due to pinning of the walls by defects, can be facilitated by exposing the material to, e.g., continuous electric-field cycling.[29] We presented evidence of such a depinning mechanism in BiFeO$_3$ ceramics in our previous study, where it was shown that the mobility of the domain walls can be increased considerably by electric-field cycling, during which the charged defects rearrange, effectively releasing domain walls.[10] Cycling experiments in that study were performed at 50 Hz, however, it is reasonable to assume that a more efficient depinning could result by application of a lower frequency field. To verify this hypothesis, we compare two polarization–electric-field hysteresis loops taken at 100 Hz and 120 kV/cm: i) before (full-line curve in Fig. 3) and ii) after (dashed-line curve in Fig. 3) applying to the BiFeO$_3$ sample three single sinusoidal waveforms with the frequency of 10, 1 and 0.1 Hz (in sequence) and amplitude of 120 kV/cm. The pinched-like shape and internal bias of the loop in Fig. 3 (full-line curve) are macroscopic manifestations of the domain-wall pinning by defects.[10] After experiencing first a field of lower frequency, the sample exhibited larger maximum polarization and more than two times larger remanent polarization (dashed-line curve in Fig. 3); the hysteresis loop is now depinched and more open. In agreement with the polarization data, a larger response was also observed in the strain (not shown). The results from Fig. 3 are therefore consistent with the assumption of a domain-wall depinning effect upon application of low-frequency field. Similar depinning by repeated switching was also observed in BiFeO$_3$ thin films.[30]

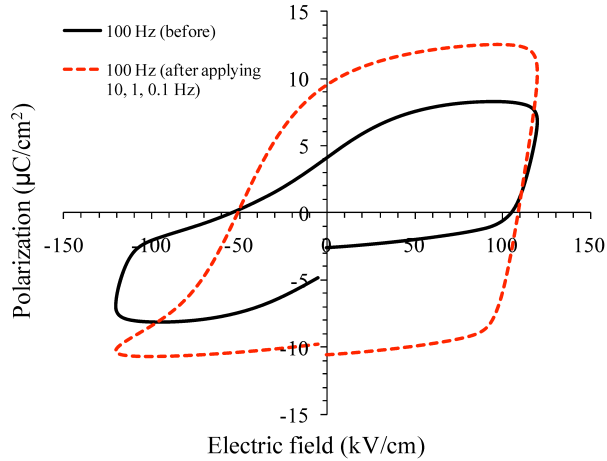

FIG. 3. Polarization–electric-field hysteresis loops of BiFeO$_3$ measured at 100 Hz and 120 kV/cm of maximum field amplitude: i) before (full-line curve) and ii) after applying sinusoidal waveforms of the same amplitude but lower frequency, i.e., 10, 1 and 0.1 Hz (dashed-line curve).

We note that polarization loops taken at frequencies of 10, 1 and 0.1 Hz showed substantial "lossy" behaviour and were not analyzed further. From this point of view, in electrically lossy materials strain–electric-field measurements can give more information on the switching process than polarization loop measurements.

In summary, large strain induced by electric field was measured in BiFeO$_3$ ceramics at low frequency (0.1 Hz) and large field amplitude (140 kV/cm). Considerable dependence of the strain on the frequency of the applied field and remanent strain suggest a possible role of non-180° domain switching mechanism. In addition, domain-wall depinning, realized at low field frequency, leads to a more efficient switching of domains and, therefore, to an increased response.

The work was carried out within the Research Program "Electronic Ceramics, Nano, 2D and 3D Structures" P2-0105 (Slovenian Research Agency).

We would like to give special thanks to Maja Hromec, Larisa Suhodolčan and Silvo Drnovšek for the preparation of the samples. Gregor Trefalt is acknowledged for help in strain measurements, Andreja Bencan-Golob for SEM and Mitja Kamplet for the grain size analysis.